\documentstyle[12pt]{article}

\newcommand{\resection}[1]{\setcounter{equation}{0}\section{#1}}

\begin{document}
\oddsidemargin 5mm
\setcounter{page}{0}
\renewcommand{\thefootnote}{\fnsymbol{footnote}}
\newpage

%\documentstyle[12pt]{article}
%\jot = 1.5ex
%\def\baselinestretch{1.65}
%\parskip 5pt plus 1pt
%\newcommand{\resection}[1]{\setcounter{equation}{0}\section{#1}}

%\catcode`\@=11

\newcommand{\beq}{\begin{equation}}
\newcommand{\eeq}{\end{equation}}
\newcommand{\bea}{\begin{eqnarray}}
\newcommand{\eea}{\end{eqnarray}}

\def\bq{\begin{quote}}
\def\eq{\end{quote}}

%\def\bbz{ Z \kern-8.9pt Z}
%\parskip 0.3cm

%       set page size
%\evensidemargin 0.0in
%\oddsidemargin 0.0in
%\topmargin -0.2in
%\textwidth 6.4in
%\textheight 8.9in
%\headsep .50in

%\topmargin -1.5cm
%\textwidth 15.5cm
%\textheight 24cm
\def\gappeq{\mathrel{ \rlap{\raise.5ex\hbox{$>$}}
{\lower.5ex\hbox{$\sim$}}}}
\def\lappeq{\mathrel{ \rlap{\raise.5ex\hbox{$<$}}
{\lower.5ex\hbox{$\sim$}}}}
\hyphenation{para-metrized}
\pagestyle{empty}
\newcommand{\dal}{\raisebox{0.085cm}
{\fbox{\rule{0cm}{0.07cm}\,}}}
\newcommand{\gef}{G_{\rm eff}}
\newcommand{\lef}{\Lambda_{\rm eff}}
\newcommand{\dt}{\partial_{\langle T\rangle}}
\newcommand{\dtbar}{\partial_{\langle\overline{T}\rangle}}
\newcommand{\al}{\alpha^{\prime}}
\newcommand{\mst}{M_{\scriptscriptstyle \!S}}
\newcommand{\mpl}{M_{\scriptscriptstyle \!P}}
\newcommand{\alg}{\alpha_{\scriptscriptstyle G}}
\newcommand{\bg}{\beta_{\scriptscriptstyle G}}
\newcommand{\bl}{\beta_{\scriptscriptstyle \Lambda}}
\newcommand{\gn}{\frac{1}{16\pi G}}
\newcommand{\gr}{G_{\!\rm eff}}
\newcommand{\eff}{\Gamma_{\!\rm eff}}
\newcommand{\lr}{\Lambda_{\rm eff}}
\newcommand{\gnn}{\frac{1}{32\pi G}}
\newcommand{\dv}{\int{\rm d}^4x\sqrt{g}}
\newcommand{\lac}{\lambda_{\scriptscriptstyle G}}
\newcommand{\act}{\widetilde{\Gamma}}
\newcommand{\lv}{\left\langle}
\newcommand{\rv}{\right\rangle}
\newcommand{\ph}{\varphi}
\newcommand{\sbar}{\,\overline{\! S}}
\newcommand{\zbar}{\bar{z}}
\newcommand{\dbar}{\,\overline{\!\partial}}
\newcommand{\tbar}{\overline{T}}
\newcommand{\ybar}{\overline{Y}}
\newcommand{\phb}{\overline{\varphi}}
\newcommand{\cm}{Commun.\ Math.\ Phys.~}
\newcommand{\pr}{Phys.\ Rev.\ D~}
\newcommand{\pl}{Phys.\ Lett.\ B~}
\newcommand{\np}{Nucl.\ Phys.\ B~}
\newcommand{\e}{{\rm e}}
\newcommand{\gsi}{\,\raisebox{-0.13cm}{$\stackrel{\textstyle
>}{\textstyle\sim}$}\,}
\newcommand{\lsi}{\,\raisebox{-0.13cm}{$\stackrel{\textstyle
<}{\textstyle\sim}$}\,}

%\begin{document}

\begin{flushright}
{CPTH-A162.0392}\\
{IC/92/51}\\
\end{flushright}
\vspace*{5mm}
\begin{center}
{\bf MODULI CORRECTIONS TO GAUGE AND GRAVITATIONAL COUPLINGS} \\
{\bf IN FOUR DIMENSIONAL SUPERSTRINGS} \\
\vspace*{2cm}
{\bf I. Antoniadis}\\
{\it Centre de Physique Th\'eorique, Ecole Polytechnique,}\\
{\it F-91128 Palaiseau, France,}\\
{\bf E. Gava$^{1,2}$}, {\bf K.S. Narain$^2$}\\
$1$ {\it Istituto Nazionale di Fisica Nucleare, sez. di Trieste, Italy,}\\
$2$ {\it International Centre for Theoretical Physics,
I-34100 Trieste, Italy}\\
\vspace*{2cm}
{\bf ABSTRACT} \\
\end{center}
We study one-loop, moduli-dependent corrections
to gauge and gravitational couplings
in supersymmetric vacua of the heterotic string. By exploiting their
relation to the integrability condition for the associated CP-odd
couplings, we derive  general expressions for them, both for $(2,2)$
and $(2,0)$ models, in terms of tree level four-point functions in
the internal $N=2$ superconformal theory. The $(2,2)$ case, in particular
symmetric orbifolds, is discussed in detail.
\vspace*{2cm}
\begin{flushleft}
%CPTH-A0??.0292\\
%IC/92/???\\
%$~$  \\
February 1992 \\
\end{flushleft}
\thispagestyle{empty}
%\mbox{}
\newpage
\setcounter{page}{1}
\pagestyle{plain}

\resection{Introduction}
Understanding the structure of the low-energy effective
Lagrangian, which governs the dynamics of the massless modes of the
heterotic superstring, is clearly an important issue. In particular
the dependence of the gauge couplings  on the moduli of
compactified space \cite{K} plays an essential role in this
context, having also implications on the problem of
Supersymmetry breaking.

This dependence has been studied extensively in the
case of the symmetric orbifold compactifications of the heterotic string
for untwisted moduli \cite{DK} and was shown to satisfy a non
renormalization theorem \cite{ANT}; namely, it is given entirely at the one
loop level and is determined by the violation of the integrability
condition of the corresponding $\Theta$-angles with respect to the
moduli. In the context of the effective field theory, this violation of
the integrability condition is related to one loop anomalous graphs
involving gauginos and matter fermions \cite{DK,DFKZ}. An expression for
these anomalies has also been obtained using effective field theory for
general heterotic vacua \cite{CARD,LOU,DF}.

In this work, starting from the full
string theory amplitude, we obtain a general formula for the violation of
the
integrability condition in terms of quantities computable from the
internal,
superconformal theory.
We also study the example of
blowing up modes for symmetric orbifolds. Finally, we use the above
method to generalize the results of ref.\cite{grav} for the
gravitational couplings.

A general four dimensional heterotic string vacuum is
obtained by tensoring four free bosons $X^{\mu}$ representing the
space-time coordinates together with their left-moving fermionic
superpartners $\psi^{\mu}$ and an internal conformal field theory with
central charge $(c,\bar{c}) = (9,22)$. $N=1$ space-time
supersymmetry implies that the internal conformal field theory must have
$N=2$ superconformal symmetry in the left-moving sector. The $N=2$
algebra involves in addition to the energy momentum tensor $T_B$, a
$U(1)$ current $J$ and a complex supercurrent $T_F^{\pm}$ carrying $U(1)$
charge $(\pm 1)$. Massless chiral or anti-chiral space-time scalars
correspond, respectively, to chiral or anti-chiral $N=2$
supermultiplets. Their lower components are primary fields $\Psi_{\pm}$
having dimensions $(\frac{1}{2},1)$ and $U(1)$ charges $(\pm 1)$ while
their upper components $\Phi_\pm$ are neutral and have dimensions $(1,1)$.
Their relevant operator product expansions (OPE) are
\begin{eqnarray}
T_F^{\pm}(z) \Psi_{\pm}(w)&=& {\rm regular} \nonumber\\
T_F^{\pm}(z) \Psi_{\mp}(w)&=& \frac{1}{z-w}\Phi_{\mp}+{\rm regular}
\nonumber\\
T_F^{\pm}(z) \Phi_{\mp}(w)&=& {\rm regular} \nonumber\\
T_F^{\pm}(z) \Phi_{\pm}(w)&=& \partial_w\bigl( \frac{\Psi_{\pm}(w)}
{z-w}\bigr)+{\rm regular}
\label{OPE}
\end{eqnarray}
\resection{$\Theta$-term and threshold corrections}
Consider a modulus field $T$ ($\tbar$) corresponding to a chiral
(anti-chiral) marginal operator. $\Theta_T$ is defined from the CP-odd
part of the on-shell three-point function of two gauge fields and the
modulus $T$. The one-loop contribution is given by:
\beq
\epsilon^{\mu\nu\lambda\rho}p_{1\lambda}p_{2\rho}\delta^{ab}\:\Theta_T ~=~
\int_{\Gamma}\frac{d^2\tau}{\tau_2}
\int\prod_{i=1}^{3} d^2z_i
\left\langle V_A^{a\mu}(p_1,z_1)\, V_A^{b\nu}(p_2,z_2)\,
V^{(-1)}_T(p_3,z_3)\, T_F(z)\right\rangle_{\rm odd}
\label{I}
\eeq
where $\tau = \tau_1 + i\tau_2$ is the Teichmuller parameter of the
world-sheet torus and  ${\Gamma}$ its fundamental domain. $V_A$'s are the
gauge vertices in zero ghost picture
\beq
V_A^{a\mu}(p,z) = :\!\bar{J}^a(\zbar )\, (\partial X^{\mu}+
ip\!\cdot\!\psi\,\psi^{\mu}) e^{ip\cdot \!X}\!:
\label{VA}
\eeq
with $\bar{J}^a$ the Kac-Moody currents. $V^{(-1)}_T$ is the $T$-vertex
operator in the $(-1)$  ghost picture
\beq
V^{(-1)}_T(p,z) = :\!\Psi_+(z,\zbar) e^{ip\cdot \!X}\!:
\label{VT}
\eeq
where the $N=1$ supercurrent insertion ($T_F=T_F^+ + T_F^- +$space-time
part) and the $(-1)$ picture for one of the vertices is due to the CP-odd
part of the amplitude, which receives contributions only from the odd spin
structure.

The four space-time zero-modes required in the odd spin structure come
from the two gauge vertices giving rise to the kinematic factor in
(\ref{I}). Furthermore, the two-point correlator of the Kac-Moody
currents is
\beq
\langle \bar{J}^a(\zbar_1 ) \bar{J}^b(\zbar_2 ) \rangle = -k \delta^{ab}
\partial_{\zbar_1}^2 \ln \bar{\theta}(\zbar_1-\zbar_2) + Q^a Q^b
\label{JJ}
\eeq
where $k$ is the level of the Kac-Moody algebra, $Q$'s are the charges of
the propagating states, and $\theta(z)$ is the Jacobi's theta-function
$\theta_1(\tau,z)$. Finally, from the OPE relations (\ref{OPE}) it
follows that the $z$-dependence in (\ref{I}) can have at most a first order
pole at $z_3$ and therefore it must be constant, due to the periodicity
of $T_F$. The $z_i$ integrations then give
\beq
\Theta_T~=~\int_{\Gamma}\frac{d^2\tau}{\tau_2} \int{d^2z}
\bar{\eta}^{-2} \langle (Q_a^2 - \pi
\frac{k_a}{\tau_2}) T_F^-(w)\Psi_+(z) \rangle_{\rm odd}
\label{II}
\eeq
where $\eta$ is the Dedekind eta function.

Now, using (\ref{OPE}) and the OPE between the $N=2$ $U(1)$ current
$J$ and the supercurrent $T_F^{mp}$, one can see that
\beq
T_F^{\mp}(w)\Psi_{\pm}(z) = \mp J(w)\Phi_{\pm}(z) \pm \oint_C
T_F^{\mp}(x)dx J(w)\psi^{\pm}(z)
\label{tps}
\eeq
where the contour $C$ encloses both $w$ and $z$. Substistuting
(\ref{tps}) in (\ref{II}), one sees that the term with the contour
integral vanishes by contour deformation, since the supercurrent
$T_F^{\pm}$ is single valued on the torus in the odd spin structure.
Thus, only the first term of the r.h.s.
of (\ref{tps}) contributes in (\ref{II}) and the result is
$$
\Theta_T~=~i\partial_T\Delta
$$
with
\beq
\Delta~=~-i
\int_{\Gamma}\frac{d^2\tau}{\tau_2}\bar{\eta}^{-2}
{\rm Tr}'_R(-1)^F F (Q_a^2 - \pi\frac{k_a}{\tau_2})
q^{L_0-\frac{c}{24}}\bar{q}^{\bar{L}_0-\frac{\bar{c}}{24}}
\label{ind}
\eeq
where $q=e^{2\pi i \tau}$ and ${\rm Tr}'$
denotes taking trace over massive states
in the internal conformal theory. Here we used the fact that the
contribution of massless states is independent of $T$ and that the
insertion of $\Phi_+$ being the marginal operator corresponding to  the
modulus $T$ just gives the the $T$-variation of $\Delta$. Repeating the
same analysis for the anti-chiral modulus $\tbar$, one obtains
$\Theta_{\tbar} = -i\partial_{\tbar}\Delta$ where the difference of sign
arises from (\ref{tps}). This shows that $\Delta$ can be identified with
the one loop gauge coupling constant \cite{DK}. It is also clear from
this discussion that non harmonicity of gauge couplings is given by the
violation of the integrability condition for the corresponding angles
$\Theta$. In the following we evaluate this violation.

As a parenthetical remark concerning (\ref{ind}), it is interesting
to notice that the quantity ${\rm Tr}(-)^F Fe^{-\beta H}$,
for the total $F=F_L-F_R$,
has been recently studied in the off-critical case in \cite{CV}, where
it is also pointed out that in the case of our interest here, conformal
and chiral, $F=F_L$, it gives a stringy generalization
of the (complex manifolds) Ray-Singer torsion.

Going back to the original expression (\ref{I}) for $\Theta_T$, its
variation with respect to $\tbar$ is obtained formally by inserting in the
three point function the vertex $V^{(0)}_{\tbar}$ in zero ghost
picture at zero momentum and  subtracting all the one-particle reducible
graphs. However, in order to regularize the short distance singularities
in the $z$-integrals, one must start with non zero momenta and take the
zero momentum limit only after performing the $z$-integrations. Indeed, one
can explicitely check in the orbifold case for untwisted moduli that
the above prescription gives the correct answer. As for the one
particle reducible graphs, the only ones that appear in the
above on-shell four point amplitude involve the antisymmetric tensor
as intermediate state, which couples to the two gauge fields
at the tree level and to the two moduli at the one loop level. These
graphs are independent of the gauge group and therefore they drop out
when one considers differences of $\frac{1}{k}\Theta_T$'s for different
gauge groups. To simplify our discussion, in the following we will
consider such differences and comment on individual gauge groups only at
the end.\\[3mm]

\resection{The integrability condition}
The variation of $\Theta_{\tbar}$ with respect to $T$ is given by the
same four point amplitude except that now $\tbar$-vertex appears in the
$(-1)$ ghost picture while the $T$-vertex in zero ghost picture.
$V^{(0)}_T$ is given by
\beq
V^{(0)}_T(p,z)=\oint dz' T_F(z') V^{(-1)}_T(p,z)
\label{VO}
\eeq
where the contour is around $z$. The violation of the integrability
condition for $\Theta$, namely
$\partial_{\tbar} \Theta_T - \partial_T
\Theta_{\tbar}~\equiv~I$, is then given by the difference
\beq
I~=~
\langle V_A V_A V^{(-1)}_T V^{(0)}_{\tbar} T_F \rangle -
\langle V_A V_A V^{(0)}_T V^{(-1)}_{\tbar} T_F \rangle
\label{D}
\eeq
Now using (\ref{VO}) in the first term of r.h.s. of (\ref{D}) we perform
the contour integral by deforming it. Due to the periodicity of $T_F$ in
the odd spin structure, the only contribution comes when the contour
encircles all other insertions. When it encircles $V^{(-1)}_T$, one gets
a contribution which cancels the second term of the r.h.s. of (\ref{D}),
while the contour integrals around $V_A$'s give total derivatives
with respect to their positions and, consequently, vanish after
$z_i$-integrations. Thus, the only term left over is the contour integral
around the $T_F$-insertion, which gives rise to the stress energy tensor
$T_B$:
\beq
I~=~\int \langle V_A(z_1) V_A(z_2) V^{(-1)}_T(z_3)
V^{(-1)}_{\tbar}(z_4) T_B(z) \rangle
\label{D1}
\eeq
where the integral is over $z_i$'s and $\tau$.

To evaluate (\ref{D1}) we extract the $z$-dependence of the above
correlator \cite{E}:
\begin{eqnarray}& &
\langle V_A(z_1) V_A(z_2) V^{(-1)}_T(z_3) V^{(-1)}_{\tbar}(z_4) T_B(z)
\rangle =~\nonumber\\[2mm] & &
-\sum_i h_i \partial_z^2 \ln\theta(z-z_i) G +
\sum_i \partial_z \ln\theta(z-z_i) \partial_{z_i}G + 2 \pi i
\partial_{\tau}G
\label{cor}
\end{eqnarray}
where $G$ is the four-point correlator
$\langle V_A(z_1) V_A(z_2) V^{(-1)}_T(z_3) V^{(-1)}_{\tbar}(z_4)
\rangle$, and $h_i$'s are the left-moving conformal weights.
Using translation invariance, one can fix one of the positions, say $z_1$,
and perform first the $z$-integral. The first term in the r.h.s. of
(\ref{cor}) is total derivative in $z$ and, therefore, upon integration
one picks up the boundary contribution  $\pi\sum_i h_i G$. The second term
is also a total derivative in $z$, but this time one has also
contributions from the branch cuts. Using the translational invariance of
$G$, and putting together the previous contribution, a straightforward
calculation yields   $$
\pi\sum_{i=2}^4 \partial_{z_i}\{[(z_i-z_1) - (\zbar_i-\zbar_1)]G\}.
$$
These terms are total derivatives with respect to $z_i$'s and the
corresponding integrations pick up only boundary contributions, which can
be identified with the $\tau$-variation of the domains of the
$z_i$-integrals of $G$. Combining with the last term of (\ref{cor}), one
finally obtains
\beq
I~=~2\pi i\int_{\Gamma}d^2\tau\partial_{\tau}
\int\prod_{i=1}^{3} d^2z_i G
\label{D2}
\eeq
The fact that the above equation involves a total derivative in $\tau$ is
not surprising. Indeed, from (\ref{D}) the integrand of $I$ is the
difference between two different choices of ghost pictures of a physical
amplitude, which is expected to be a total derivative. For non zero
external momenta $I$ would have vanished after integration over
$\tau$. However, as we will see below, setting the momenta to zero
before the $\tau$-integration gives a non trivial result.

The gauge vertices provide the four space-time fermion zero modes
required in the odd spin structure giving rise to the kinematic factor in
(\ref{I}). Moreover the Kac-Moody currents are replaced by their zero
modes as we are considering differences of gauge groups. Since $T$ and
$\tbar$ are marginal operators their vertices have no first order pole
in the anti-holomorphic sector (i.e. bosonic sector). Therefore we can
set the external momenta to zero and the short distance singularities can
be handled by angular integration.  Integration over $z_1$ and $z_2$ then
gives two powers of $\tau_2$ which cancel the powers of $\tau_2$ coming
from the integration over the loop momentum. One is then left with $\tau$
derivative of the two point function of $T$ and $\tbar$ in the internal
superconformal theory:
\beq
I\sim\int d^2\tau{\bar{\eta}}^{-2}\partial_{\tau}\int d^2 z\langle
V^{(-1)}_T
(z)V^{(-1)}(0)_{\tbar}\rangle_{\rm odd}
\label{D21}
\eeq
$\tau$ integration then gives only the boundary
term, i.e. $\tau_2 \rightarrow \infty$ limit, giving rise to a sum over
four point functions on the sphere:
\beq
I \sim \displaystyle\lim_{\tau_2\rightarrow
\infty}\int d\tau_1{\bar{\eta}}^{-2}{\rm Tr}(-1)^F Q^2
q^{L_0-\frac{c}{24}}\bar{q}^{\bar{L}_0-\frac{\bar{c}}{24}} \int d^2 x ~
x^{-\frac{1}{2}}\langle V(0)  V^{(-1)}_T(x) V^{(-1)}_{\tbar}(1)
\bar{V}(\infty) \rangle
\label{D3}
\eeq
where the
trace is over all states of the internal conformal theory in the Ramond
sector with vertices $V$ and $\bar{V}$
(conjugate vertices) in the $(-\frac{1}{2})$ ghost picture. The
integration of $x$ is in an annulus within radii $|q|^{\frac{1}{2}}$ and
$|q|^{-\frac{1}{2}}$ and the factor  $x^{-\frac{1}{2}}$ comes from the
transformation of the torus coordinates to the annulus coordinates (recall
that $V^{(-1)}_T$ has dimension $(\frac{1}{2},1)$). If the $x$ integration
is regular then it is clear from (\ref{D3}) that only the trace over
massless states contribute. The only possibility for a massive state to
contribute in this limit is if the behavior of $x$ integral near the
boundaries of annulus diverge. Consider for instance a massive state,
i.e. with $L_0>\frac{3}{8}$ and $\bar{L}_0>1$. The behaviour of the
correlation function as $x$ goes to zero is $x^{-L_0-\frac{1}{2}+h}
\bar{x}^{-\bar{L}_0-1+\bar{h}}$ where $(h,\bar{h})$ are the conformal
weights of the intermediate state. The angular integration of $x$ in
(\ref{D3}) is non zero only if  $-L_0-1+h = -\bar{L}_0-1+\bar{h}$. If
$h=L_0$, the $x$ integration has only logarithmic divergence in $|q|$.
The integration of $\tau_1$ enforces the level maching condition and the
result vanishes due to $|q|^{m^2}$ suppression, with $m$ being the mass.
If $h\ne L_0$, the $x$ integral after including the remaining terms in
(\ref{D3}) behaves as
\beq
\frac{1}{L_0-h}
q^{{\frac{1}{2}}(L_0+h-{\frac{3}{4}})}\bar{q}^{{\frac{1}{2}}
(\bar{L}_0+\bar{h}-2)}
\label{L0}
\eeq
Note that as the trace is over Ramond sector
both $L_0$ and $h$ are greater than or equal to $\frac{3}{8}$. Performing
the $\tau_1$ integration and taking the limit $\tau_2$ to infinity, once
again the result vanishes due to the mass suppression, assuming that
there is no accumulation point of masses at zero which should be the case
for compact internal spaces. Therefore, at generic points in moduli space
only massless states contribute in (\ref{D3}).

The only subtlety arises if there are special subspaces of moduli space
where there are extra massless states. Away from these subspaces, these
extra states do not contribute to (\ref{D3}). However as one approaches
these regions of moduli spaces, the masses of these extra particles
become arbitrarily small. In particular, if one sits at a point on this
subspace, the extra massless states do contribute to (\ref{D3}).
For example, if $T$, $\tbar$ correspond to vector fields which are tangent
to this subspace, then the contribution of the extra massless states is
finite
just on it, and vanishes away from it due to the limit
$|q|^{m^2}\rightarrow
0$ as $|q|\rightarrow 0$. This gives rise to a discontinuity in $\Delta_
{T\tbar}$. On the other hand, if one considers moduli $T$, $\tbar$ that are
normal to the subspace, then the contribution of these states gives rise
to a $\log|q|$ divergence, as it appears from (\ref{L0}). This is a problem
which arises also in field theory when doing mass perturbation expansion
around zero mass.
A possible way to handle this problem is to
introduce an infrared cutoff $\Lambda$ in the $\tau_{2}$ integration; the
result will then be continuous and duality invariant. The above discussion
applies also to the case of $x$ near infinity.

In the following we will ignore this subtleties and consider the case
where there is no divergence.
Thus (\ref{D3}) becomes
\beq
I \sim {\rm Tr}(-1)^F Q^2 \int d^2 x ~
x^{-\frac{1}{2}}\langle V^R(0)  V^{(-1)}_T(x) V^{(-1)}_{\tbar}(1)
\bar{V}^R(\infty) \rangle
\label{D4}
\eeq
where the trace is only over massless states in the Ramond sector.

Now we will relate the above four point amplitude to that of a physical
4-boson amplitude. Consider first a physical amplitude involving four
scalars
\beq
A~=~\langle V_{-}^{(-1)}(p_1,z_1)  V^{(0)}_{+}(p_2,z_2)
V^{(0)}_{-}(p_3,z_3) V_{+}^{(-1)}(p_4,z_4) \rangle
\label{A}
\eeq
where the subscripts $+$ and $-$ refer to the $N=2$ chirality and
$$
V_{\pm}^{(-1)}(p) = \Psi_{\pm}e^{ip\cdot X}
$$
$$
V_{\pm}^{(0)}(p) = (\Phi_{\pm}+ip\cdot \psi \Psi_{\pm})e^{ip\cdot X}
$$
Using the above in (\ref{A}), we obtain
\begin{eqnarray}
A ~&=&~\langle  V_{-}^{(-1)}(p_1,z_1) \Phi_{+}e^{ip_2\cdot X}(z_2)
\oint dz
\frac{z-z_1}{z_3-z_1}T^{+}(z)V_{-}^{(-1)}(p_3,z_3)V_{+}^{(-1)}(p_4,z_4)
\rangle \nonumber\\
&-&\frac{p_2\cdot p_3}{z_2-z_3}\langle  V_{-}^{(-1)}(p_1,z_1)
V_{+}^{(-1)}(p_2,z_2) V_{-}^{(-1)}(p_3,z_3)V_{+}^{(-1)}(p_4,z_4)\rangle
\label{A1}
\end{eqnarray}
Now pulling the contour around and noting that there is no singularity at
infinity we see that the only contribution comes from the contour
integral around $z_2$. The result is
\beq
A~=~-\frac{p_2\cdot p_4}{z_3-z_1}\frac{z_4-z_1}{z_4-z_2}
\langle  V_{-}^{(-1)}(p_1,z_1) V_{+}^{(-1)}(p_2,z_2) V_{-}^{(-1)}(p_3,z_3)
V_{+}^{(-1)}(p_4,z_4)\rangle
\label{A2}
\eeq

To relate the amplitude $A$ to that of two fermions and two bosons, we
use the space-time supersymmetry relation
\beq
\Psi_{\pm}(w) = \oint dz (z-w)^{-\frac{3}{4}} e^{\pm i
\frac{\sqrt{3}}{2}H(z)} \Psi_{\pm}^R (w)
\label{R}
\eeq
where $\Psi_{\pm}^R$ is the Ramond vertex of the internal
$N=2$ superconformal theory  with $U(1)$ charges $\mp \frac{1}{2}$, and $H$
bosonises the corresponding $U(1)$ current $\sqrt{3}\partial H$. Note
that  $e^{\pm i\frac{\sqrt{3}}{2}H}$ is just the internal part of the
space-time supersymmetry current. Using (\ref{R}) we can replace the
bosonic vertices at $z_1$ and $z_4$ with two contour integrals $\oint dz$
and $\oint dw$, respectively, around Ramond vertices. The $z$ and $w$
dependence can be extracted explicitely from the $U(1)$ charges of
all vertices with the result
\begin{eqnarray}
& &\langle  V_{-}^{(-1)}(p_1,z_1) V_{+}^{(-1)}(p_2,z_2)
V_{-}^{(-1)}(p_3,z_3) V_{+}^{(-1)}(p_4,z_4)\rangle = \nonumber\\
& &\oint dz \oint dw \frac{(z-z_4)^{\frac{1}{4}}(w-z_1)^{\frac{1}{4}}
(z-w)^{-\frac{3}{4}}}{(z-z_1)(w-z_4)}\bigl(\frac{z-z_3}{z-z_2}\bigr)^
{\frac{1}{2}}\bigl(\frac{w-z_2}{w-z_3}\bigr)^{\frac{1}{2}}\nonumber\\
& &\langle  V_{-}^R(p_1,z_1) V_{+}^{(-1)}(p_2,z_2) V_{-}^{(-1)}(p_3,z_3)
V_{+}^R(p_4,z_4)\rangle
\end{eqnarray}
where $V_{\pm}^R(p)=\Psi_{\pm}^R e^{ip\cdot X}$.  Substituting
the above equation in (\ref{A2}), performing the contour integrals and
fixing $SL(2,C)$ invariance by setting $z_1=0$, $z_2=x$, $z_3=1$ and
$z_4=\infty$ we obtain
\begin{eqnarray}
& &x^{-\frac{1}{2}}\langle  V_{-}^R(p_1,0) V_{+}^{(-1)}(p_2,x)
V_{-}^{(-1)}(p_3,1) V_{+}^R(p_4,\infty)\rangle ~=~\nonumber\\
& &-\frac{1}{u}\langle V_{-}^{(-1)}(p_1,0)  V^{(0)}_{+}(p_2,x)
V^{(0)}_{-}(p_3,1) V_{+}^{(-1)}(p_4,\infty) \rangle
\label{A3}
\end{eqnarray}
where $u=p_2\cdot p_4$. Similarly, by exchanging the role of $0$ and
$\infty$ one gets
\begin{eqnarray}
& &x^{-\frac{1}{2}}\langle  V_{+}^R(p_1,0) V_{+}^{(-1)}(p_2,x)
V_{-}^{(-1)}(p_3,1) V_{-}^R(p_4,\infty)\rangle ~=~\nonumber\\
& &\frac{1}{s}\langle V_{+}^{(-1)}(p_1,0)  V^{(0)}_{+}(p_2,x)
V^{(0)}_{-}(p_3,1) V_{-}^{(-1)}(p_4,\infty) \rangle
\label{A4}
\end{eqnarray}
where $s=p_2\cdot p_1$ and the relative sign is due to the exchange of
space-time fermions. Note that the terms appearing on the l.h.s. of the
above two equations give precisely the contribution of matter
fields in (\ref{D4}).

The terms appearing in the r.h.s. of (\ref{A3}) and (\ref{A4}) are four
boson tree-level amplitudes and they can be calculated at the level of
the effective field theory. Since we are interested in the zero-momentum
limit, only the terms involving two derivatives are relevant and the
result is \cite{DKL}:
\beq
A\bigl( C(p_1) T(p_2) \tbar(p_3) \overline{C}(p_4)\bigr) ~=~
s(R_{C\overline{C}T\tbar} + \frac{u}{t} G_{C\overline{C}} g_{T\tbar})
+ V_{C\overline{C}T\tbar}
\label{A5}
\eeq
where $C$ ($\overline{C}$) are chiral (anti-chiral) matter fields (not
necessarily complex conjugate), $G_{C\overline{C}}=\partial_C
\partial_{\overline{C}}K$ and $g_{T\tbar}=\partial_T \partial_{\tbar}K$
are the matter and moduli metrics, respectively, with $K$ being the Kahler
potential and $R_{C\overline{C}T\tbar}$ is the corresponding Riemann
tensor. $V_{C\overline{C}T\tbar}$ denotes the potential contribution
which generate a mass to the fields $C$, $\overline{C}$ from the $T$
expectation value. If this term is non zero, it induces a divergence in
(\ref{D4}) via (\ref{A3}) and (\ref{A4}), as also discussed before.
As also previously stated we will not consider such a situation.
The second
term in the r.h.s. of (\ref{A5}) accounts for graviton exchange in the
$t$-channel. When there is enhanced gauge symmetry, there is an additional
contribution in the $t$-channel due to the exchange of gauge bosons which
couple simultaneously to $T$, $\tbar$ and $C$, $\overline{C}$ fields.
However this term is proportional to ${\rm Tr}_C Q'_C$, where $Q'$ is the
charge with respect to the enhanced gauge symmetry, and it vanishes unless
there is an anomalous $U(1)$ factor. In the following we restrict
ourselves to moduli which are neutral under such anomalous  $U(1)$
factors.

Similarly, (\ref{A3}) involves
$A\bigl( \overline{C}(p_1) T(p_2) \tbar(p_3) C(p_4)\bigr)$
which is obtained
from (\ref{A5}) by interchanging $s$ and $u$. Substituting these
expressions in (\ref{A4}) and (\ref{A3}) one has to consider their
difference in order to take into account $(-1)^F$ in (\ref{D4}).
Moreover, the amplitude (\ref{D4}) obtained from the degeneration of the
torus is a forward amplitude and therefore the correct zero-momentum
limit is to first take $t\rightarrow 0$ and then set the momenta equal
to zero. The final expression is
\beq
I_{\rm matter}~=~\sum_{C,\overline{C}}{\cal{T}}(C)G^{C\overline{C}}
(2R_{C\overline{C}T\tbar} - G_{C\overline{C}} g_{T\tbar})
\label{Dmat}
\eeq
where $I_{\rm matter}$ is the contribution of matter fields to $I$
and ${\cal{T}}(C)$ is the index of the representation of the $C$ field.
The remaining contribution to the trace in (\ref{D4}) is due to the
gauginos, which can be explicitely calculated using the fact that the
internal part of their vertices is $e^{\pm i \frac{\sqrt{3}}{2}
H}\bar{J}$, $\bar{J}$ being the Kac-Moody current. The answer is
\beq
I_{\rm gauge}~=~{\cal{T}}({\rm Ad})g_{T\tbar}
\label{Dga}
\eeq
where Ad denotes adjoint representation.\\[3mm]
\resection{$(2,2)$ models and orbifolds}
In the case of symmetric $(2,2)$ compactifications, the gauge group is
$E_6 \times E_8$ and the matter fields transform as $27$ or
$\overline{27}$ under $E_6$ and they are in one-to-one correspondence with
the moduli : $27$'s are related to $(1,1)$ moduli and $\overline{27}$'s to
$(1,2)$ moduli.  Furthermore, the moduli metric is block-diagonal with
respect to these two types of moduli. In the following $(1,1)$ and
$(1,2)$ moduli will be collectively denoted by $T_1$ and $T_2$ and their
metrics by $g^1$ and $g^2$. Equation (\ref{Dmat}) is then further
simplified due to relations among various four-point string amplitudes,
emerging from the right moving $N=2$ internal supersymmetry \cite{DKL}.
In fact, relating amplitudes involving two matter and two moduli fields
with those involving four moduli, one obtains for $(1,1)$ moduli
\beq
G^{C\overline{C}}R_{C\overline{C}T_1\tbar_1}~=~ R^1_{T_1\tbar_1} -
[\frac{1}{3}(h_{(1,1)}-h_{(1,2)})]g^1_{T_1\tbar_1} \label{rel}
\eeq
where $h_{(1,1)}$ and $h_{(1,2)}$ denote the number of $(1,1)$ and
$(1,2)$ moduli, respectively, and $R_1$ is the Ricci-tensor constructed
with the metric $g_1$,
$R^1_{T\tbar}=\partial_T\partial_{\tbar}\ln \det g_1$. A similar
expression is obtained for $(1,2)$ moduli by interchanging
$(1\leftrightarrow 2)$ and $h_{(1,1)} \leftrightarrow h_{(1,2)}$.
Substituting (\ref{rel}) in (\ref{Dmat}) and adding (\ref{Dga}) one gets
\begin{eqnarray}
I_1(E_6)&=&\bigl[{\cal{T}}(E_6) - {\cal{T}}(27)
(\frac{5}{3} h_{(1,1)}+\frac{1}{3}h_{(1,2)})\bigr]g^1 + 2{\cal{T}}(27)R^1
\nonumber\\
I_1(E_8)&=&{\cal{T}}(E_8)g^1
\label{DN2}
\end{eqnarray}
Here ${\cal{T}}(27)=3$, ${\cal{T}}(E_6)=12$ and ${\cal{T}}(E_8)=30$.
A similar expression is obtained for $I_2$ by interchanging
$(1\leftrightarrow 2)$ and $h_{(1,1)}\leftrightarrow h_{(1,2)}$. Equation
(\ref{DN2}) is identical with that obtained in \cite{DF} using effective
field theory anomalous graphs. Note that the result (\ref{DN2}) continues
to hold when there is an enhanced gauge symmetry. Although the
relationship between the four-moduli Riemann tensor and the one with two
matter and two moduli indices gets modified by additional terms due to the
exchange of additional gauge bosons, these terms being proportional to
$Q'_C$ (additional gauge charges of $C$'s) drop out after taking the
trace in order to obtain (\ref{rel}).
The above expressions can be further related to Yukawa
couplings by using the relation \cite{DKL,CREM}
\beq
R^1_{T\tbar}~=~(h_{(1,1)}+1)g^1_{T\tbar}-e^{2K_1}|W^1|^2_{T\tbar}
\label{Ric}
\eeq
where $W^1$ and $W^2$ are holomorphic functions of the moduli fields and
give the Yukawa couplings of $27$ and ${\overline{27}}$ families,
respectively. A similar relation is valid for $R^2$ by interchanging
$(1\leftrightarrow 2)$ and $h_{(1,1)} \leftrightarrow h_{(1,2)}$.

We now apply the above general results to the case of $(2,2)$ orbifolds.
Consider first the untwisted moduli. For simplicity, we discuss the case
when the three internal planes are orthogonal to each other and obtain the
dependence of the gauge couplings on the moduli associated to these three
planes. The dependence on the remaining untwisted moduli can then be
obtained using the full duality group. Let $T$ denote the $(1,1)$ modulus
associated with the first plane; its vertex at zero-momentum being
$\partial X_1 \dbar \bar{X}_1$, where $X_1$ is the complex
coordinate in the plane normalized so that the double pole in the two
point function $\langle \partial X_1 \partial \bar{X}_1\rangle$
comes with the coefficient $\frac{1}{T+\tbar}$. The moduli metric is
then given by $g_{T\tbar}=(T+\tbar)^{-2}$. To compute the Ricci tensor
entering in (\ref{DN2}) we also need the $T$-dependence of the remaining
components of the moduli metric. The components of the metric along the
moduli associated to the other two planes are independent of $T$. However,
the untwisted moduli related to the angles between the first and the other
two planes with vertices being of the form
$\partial X_1 \dbar \bar{X}_{2,3}$, lead to a metric proportional
to $(T+\tbar)^{-1}$. Let $n_1$ be the number of such untwisted moduli.
As for the twisted moduli (blowing-up modes), only the $(1,1)$ type which
carry non zero charge with respect to the $U(1)$ current
\footnote{Recall that in the case of $(2,2)$ orbifolds there is an
enhanced gauge symmetry : besides the $SO(10)$ there are three Cartan
generators  $\bar{J}_i=\bar{\psi_i}\bar{\psi_i}^*$,
$i=1,2,3$, corresponding to the three planes. Of these, the diagonal
generator is the $N=2$ $U(1)$ current which together with the $SO(10)$
gives rise to $E_6$.}
$\bar{J}_1=\bar{\psi_1}\bar{\psi_1}^*$ have a $T$-dependent
metric proportional to $(T+\tbar)^{-1}$ \cite{DKL}. $\bar{\psi_1}$
is the right-moving $N=2$ superpartner of the coordinate $X_1$. Let $b_1$
be the number of such twisted $(1,1)$ moduli. Then, the Ricci-tensor is
$$
R^1_{T\tbar}=(2+n_1+b_1)g^1_{T\tbar}
$$
Substituting this expression in (\ref{DN2}) and taking the difference
between $E_6$ and $E_8$ contributions, one gets
\beq
I_1(E_6)-I_1(E_8)~=~[\frac{1}{3}(h_{(1,1)}-h_{(1,2)})-4-2N_1]{\cal{T}}(27)
\frac{1}{(T+\tbar)^2}
\label{orb}
\eeq
where $N_1$ is the total number of $(1,1)$ moduli which have no singular
OPE with $T$ and $\tbar$. One can explicitly check in examples that if
$T$ corresponds to a plane which is twisted under all non trivial
elements of the orbifold group the r.h.s. of (\ref{orb}) vanishes, in
agreement with known results \cite{DK}. In the case when some non
trivial element of the orbifold group leaves this plane untwisted, one
gets a non vanishing coefficient related to the $\beta$-function of the
corresponding $N=2$ subsector. More precisely, in this case one has
\beq
I_1(E_8)-I_1(E_6)~=~\frac{2}{\rm ind}(\hat {b}_8-\hat {b}_6)\frac {1}
{(T+\tbar)^2}
\eeq
in which ind is the index of the little subgroup of the plane in the
full orbifold group, and $\hat{b}_8$, $\hat{b}_6$ are the $\beta$-function
coefficients of the corresponding $N=2$ subsectors.

In the case of blowing-up modes $(B,\overline{B})$, one can calculate the
dependence of the gauge couplings on these moduli in a perturbative
expansion. For instance, the first non trivial term proportional to
$B\overline{B}$ is just given by the violation of the integrability
condition (\ref{DN2}) for $B$, $\overline{B}$. From the initial expression
(\ref{D1}) this involves a correlation function on the torus involving
$B$ and $\overline{B}$, which can be non zero only if $B$ and
$\overline{B}$ come
from opposite twisted sectors and fixed points. $\overline{B}$ is then the
complex conjugate of $B$. The untwisted moduli dependence of the metric
$g_{B\overline{B}}$ has already been discussed in the previous paragraph,
and, on
the other hand, the corresponding Ricci tensor can be calculated using
(\ref{Ric}), so that one has:
\beq
\Delta_{E_6}~=~\frac{2}{\rm ind}\hat{b}_6 \log(|\eta(iT)|^4(T+\tbar))
+B\overline{B}\{g^1_{B\overline{B}}[{\cal{T}}(E_6)+{\cal{T}}(27)({\frac
{\chi}{3}}+2)]-2{\cal{T}}(27)e^{2K_1}|W^1|^2_{B\overline B}\}
\label{twist}
\eeq
where $\chi=h_{(1,1)}-h_{(1,2)}$ is the Euler number,
and the Yukawa couplings are given for instance in \cite{Yuk}.\\[3mm]
\resection{Gravitational couplings}
We now discuss the moduli dependence of gravitational couplings,
generalizing the results of ref\cite{grav} which were obtained for
orbifolds. In this case, the role of the gauge couplings and the
$\Theta$-angles is played by the coefficients of the Gauss-Bonnet
combination and the $R\tilde{R}$ term, respectively. The study of the
moduli dependence of this coupling involves the calculation of the
violation of the integrability condition for $\Theta^{\rm grav}$. As
before, we start from the CP-odd part of the on-shell three-point
function of two gravitons and one modulus. Equation (\ref{I}) is now
replaced by
\beq
\epsilon^{\mu\nu\lambda\rho}p_{1\lambda}p_{2\rho}p_2^{\alpha}p_1^{\beta}
\:\Theta_T^{\rm grav}~=~
\int_{\Gamma}\frac{d^2\tau}{\tau_2}
\int\prod_{i=1}^{3} d^2z_i
\left\langle V_h^{\alpha\mu}(p_1,z_1)\, V_h^{\beta\nu}(p_2,z_2)\,
V^{(-1)}_T(p_3,z_3)\, T_F(z)\right\rangle_{\rm odd}
\label{Ig}
\eeq
where $V_h$'s are the graviton vertices obtained from (\ref{VA}) by
replacing $\bar{J}^a$ with $\dbar X^{\alpha}$. The four space-time
zero-modes are provided by the graviton vertices, giving rise to the
kinematic factor $\epsilon^{\mu\nu\lambda\rho}p_{1\lambda}p_{2\rho}$.
The remaining space-time part of the correlator (\ref{Ig}) is
\beq
\int d^2z_1\int d^2z_2 \langle \dbar X^{\alpha}e^{ip_1\cdot \!X}(z_1)
\dbar X^{\beta}e^{ip_2\cdot \!X}(z_2)e^{ip_3\cdot \!X}(z_3)\rangle
\label{bosg}
\eeq
as only the internal part of $T_F$ contributes in (\ref{Ig}). Of all
possible ways of contractions, the only non vanishing one is when
$\dbar X^{\alpha}$ is contracted with $e^{ip_2\cdot \!X}$ and
$\dbar X^{\beta}$ with $e^{ip_1\cdot \!X}$ : the remaining ones are total
derivatives (on-shell) in $\zbar_1$ or $\zbar_2$ and, consequently,
vanish upon integration. Then, (\ref{bosg}) provides the kinematic factor
$p_2^{\alpha}p_1^{\beta}$ multiplying \cite{grav}
\beq
\tau_2\int d^2z_1\langle\dbar X(z_1)X(0)\rangle^2~=~
-2\pi i \tau_2\partial_{\bar{\tau}}\bigl( \frac{1}{\tau_2
\bar{\eta}^2}\bigr)
\label{corg}
\eeq

The internal part of (\ref{Ig}) being the same as in the gauge case, one
obtains for $\Theta_T^{\rm grav}$ an expression identical to (\ref{II})
after replacing $Q_a^2$ with
\beq
Q_{\rm grav}^2\equiv -\partial_{\bar{\tau}}\ln (\bar{\eta}^2)
\label{Qgr}
\eeq
The rest of the analysis can now be repeated exactly as for the gauge
couplings leading to (\ref{ind}) for $\Delta^{\rm grav}$ after replacing
$Q_a^2$ with $Q_{\rm grav}^2$. Furthermore, considering the differences
$\Theta_T^{\rm grav}-\frac{1}{k}\Theta_T^{\rm gauge}$ one finds equation
(\ref{D3}) for the violation of the integrability condition, and once
again only the massless states contribute in the trace. There are two
possible ways to get these massless states:
\begin{itemize}
\item[({\it a\/})]
{}From an excitation of the space-time part, i.e. gravitinos and dilatino
corresponding to the first term of the expansion of $\bar{\eta}^2$ in
(\ref{corg}), giving rise to $Q_{\rm grav}^2=\frac{11}{6}$.
\item[({\it b\/})]
{}From the lowest order term in (\ref{corg}) corresponding to gauginos and
matter fermions with $Q_{\rm grav}^2=-\frac{1}{12}$.
\end{itemize}
Equation (\ref{D4}) then follows with the above values of $Q_{\rm
grav}^2$. The contribution of gravitinos, dilatino and gauginos can be
determined by a direct calculation as in the case of the adjoint
representation for gauge couplings with the result
\beq
I^{\rm grav}_V~=~(\frac{11}{6}-\frac{1}{12}n_V)g_{T\tbar}
\label{Dgrv}
\eeq
where $n_V$ is the number of massless vectors. The contribution of matter
fields can also be calculated by the same steps as in the gauge case,
resulting
\beq
I^{\rm
grav}_S~=~-\frac{1}{12}(2\sum_{a,\overline{a}}G^{a\overline{a}}
R_{a\overline{a}T\tbar} - n_S g_{T\tbar})
\label{Dgrs}
\eeq
where the sum is over all massless complex scalar fields (excluding
dilaton), $G_{a\overline{a}}$ being the corresponding metric and $n_s$
their number.

In the case of $(2,2)$ models, (\ref{Dgrv}) and (\ref{Dgrs}) together
become (for example for $(1,1)$ moduli)
\beq
I^{\rm grav}_1~=~[\frac{11}{6}+\frac{1}{12}(n_S-n_V)+
\frac{3}{2}(h_{(1,1)}-h_{(1,2)})]g^1_{T\tbar} - \frac{14}{3}R^1_{T\tbar}
- \frac{1}{6}{\sum_{a,\overline{a}}}'G^{a\overline{a}}
R_{a\overline{a}T\tbar}
\label{22gr}
\eeq
where the sum is restricted to the $E_6\times E_8$ singlets which are not
$(1,1)$ or $(1,2)$ moduli. Unfortunately, this expression cannot be
further simplified due to the present lack of undestanding of these extra
massless singlet states. Of course for specific models such as orbifolds
and Gepner models these quantities can be computed.\\[3mm]
\resection{Examples and conclusions}
For instance, consider $(2,2)$ orbifolds, where the three internal
planes are orthogonal to each other. Associated to them there are
$U(1)$ charges $F^i$ in the fermionic sector ( the $N=2$, $U(1)$
charge being the sum of the $F^i$'s ). Let then $T$ denote the
$(1,1)$ modulus corresponding to the first plane: the right moving
part of the vertex operator for it is then ${\dbar}X_1$ (${\dbar}
\overline{X}_1$) where $X_1$, $\overline{X}_1$ are complex
coordinates on the first plane which diagonalize the orbifold group
action: $g: X_1 \rightarrow e^{2\pi i{\alpha}_1}X_1$ , $\alpha_1$
being a rational number $0\leq \alpha <1$. Tha non-moduli singlet
states are generally constructed by applying certain numbers of
right moving oscillator modes on the twist fields. Let $n_i$ ($
-\overline{n}_i$) be the numbers of ${\dbar}X_i$ (${\dbar}
\overline{X}_i$) oscillator modes in a given state.

The contribution to the curvature term in (\ref{22gr}) from the non-moduli
singlets, from the sector where the first plane is twisted can be
calculated directly from the 4-point function and is:
\beq
-\frac{1}{6}\{\sum[(1-\alpha_1)+n_1]-b_1\},
\label{orb1}
\eeq
where $b_1$ are the twisted moduli that have singular OPE with $T$, and
the sum is over chiral states ( $\sum F_i=1$ ) in these sectors.
Similarly the contribution from the sectors that leave $X_1$ untwisted
( i.e. $\alpha_1=0$ ) is given by:
\beq
-\frac{1}{6}\sum F_1,
\label{orb2}
\eeq
where the sum is over the chiral states in such
sectors.

For specific orbifold models one can get all the above numbers from
the spectrum. One can check that for models with no $N=2$ subsector
( such as $Z_3$ and $Z_7$ orbifolds ), $I_{1}^{grav}+2I_{E_8}=0$
( note that this is the correct modular invariant combination ).
On the other hand for models with $N=2$ subsector one obtains:
\beq
I_{1}^{grav}+2I_{E_8}~=~\frac{2}{\rm ind}(\frac{1}{2}
\hat{b}_{grav}+2\hat{b}_8)\frac{1}{(T+\tbar)^2},
\label{orb3}
\eeq
where as before ind is the index of the little group of the plane,
$\hat{b}_{grav}$ and $\hat{b}_8$ are the trace anomaly and $\beta$-
funtion coefficients for the corresponding $N=2$ space-time
supersymmetric theory obtained when the orbifold is taken to be
the above little group.

For example, take the $Z_4$ orbifold; the orbifold group is generated
by an element $g$ with $(\alpha_1,\alpha_2,\alpha_3)=(1/4,1/4,1/2)$.
The gauge group is $E_{8}\times E_{6}\times SU(2)\times U(1)$ and the
left-moving chiral $E_6$ singlets are shown below, with the relevant
quantum numbers $(F_1,F_2,F_3)$ and $(n_1,n_2,n_3)$, for untwisted and
twisted sectors.
{}From the untwisted sector: $(F_1,F_2,F_3)=(1,0,0)$,
or $(0,1,0)$, $SU(2)$ doublets.\\
{}From the $g$-twisted sector: $(F_1,F_2,F_3)=(1/4,1/4,1/2)$ and
for $(n_1,n_2,n_3)$ one has the following possibilities: $16(2,0,0)$,
$16(1,1,0)$,$16(0,2,0)$,$16(0,0,1)$,$16(0,0,-1)$,all singlets, and
$16(1,0,0)$,
$16(0,1,0)$,which are doublets.\\
{}From the $g^2$-twisted sector: $(F_1,F_2,F_3)=(1/2,1/2,0)$ one has for
$(n_1,n_2,n_3)$ the following cases: $10(1,0,0)$,$10(0,1,0)$,$6(-1,0,0)$,
$6(0,-1,0)$, which are doublets, and $16(0,0,0)$ singlets.\\
Plugging these values in (\ref{orb1}) and (\ref{orb2})
for the case where $T$ is the
$(1,1)$ modulus corresponding to the first plane, one finds $I_1^{grav}+
2I_8=0$, as expected because this plane is twisted by all the nontrivial
elements of the orbifold group. On the other hand, if $T$ corresponds
to the $(1,1)$ modulus of the third plane, one finds: $I_1^{grav}+2I_8=
\frac{1}{2}\hat{b}_{grav}+2\hat{b}_8$, consistent with (\ref{orb3}) as the
index of the little group of the third plane is $2$. Here $\hat{b}_{grav}$
and $\hat{b}_8$ are the trace anomaly and $E_8$ $\beta$-function for the
$N=2$ spacetime SUSY theory corresponding to the $Z_2$ orbifold generated
by $g^2$.

Although we have discussed here the examples of $(2,2)$ orbifolds, one
can easily generalize this to $(2,0)$ orbifolds. The only difference
is that now that the relation (\ref{rel}) between the mixed moduli-matter
and moduli Riemann tensors is no longer valid. However one can use
relations
similar to (\ref{orb1}) and (\ref{orb2}) to evaluate the contribution
of matter fields.

To conclude, the main result here is that the 1-loop violation of the
integrability condition for $\Theta$-angles (gauge and gravitation) and
consequently non-harmonicity of the gauge and gravitational couplings
is expressed in terms of tree level four-point amplitudes. For $(2,2)$
models this quantity for the gauge coupling case, was further related
to the metric on the moduli space due to relations among various four
point amplitudes.

One of the issues which needs further study is that of special subspaces
of the Moduli Space where one has extra massless states. As discussed in
in section $2$, if one evaluates
threshold corrections at a generic point (i.e. away from these subspaces)
these extra massless states do not contribute, however on these special
subspaces they do contribute. This could introduce a discontinuity in the
threshold correction as function of moduli. For example, many of the
massless
singlets in orbifold backgrounds become massive when the blowing-up modes
are turned on; consequently if one approaches the orbifold point from
smooth manifolds the result would be different from the one obtained
at the orbifold point. As suggested also in section $2$,
a possible starting point to analyze this problem
could be to introduce an infrared cut-off (e.g. a cut-off in the $\tau_2$
integration); the resulting threshold corrections would then be continuous
and duality invariant in moduli. A full understanding of this issue
requires
however further study.

\vskip 0.5cm
\noindent
{\bf Acknowledgements}
I.A. and K.S.N. thank respectively ICTP and Ecole Polytechnique
for hospitality during the completion of this work. I.A. thanks also
S.Ferrara and T.Taylor for discussions.
\vskip 0.5cm
%\newpage

\end{document}